\documentclass[format=acmsmall, authorversion=true, screen=true]{acmart}

\acmJournal{TOIS}
\acmVolume{1}
\acmNumber{1}
\acmArticle{1}
\acmYear{2020}
\acmMonth{1}
\copyrightyear{2021}
\setcopyright{acmlicensed}
\acmJournal{TOIS}
\acmPrice{15.00}\acmDOI{10.1145/xxxxxxx}

\usepackage[utf8]{inputenc}
\usepackage{graphicx}
\usepackage{subcaption}
\usepackage{multirow}
\usepackage{bbm}
\usepackage{hyperref}
\usepackage{pgfplots}
\usepackage{accsupp}
\usepackage{pifont}
\usepackage[para]{footmisc}
\usepackage[show]{chato-notes}

\def\addlegendimage{\csname pgfplots@addlegendimage\endcsname}
\pgfplotsset{
cycle list={%
{draw=black,mark=star,solid},
{draw=black, mark=square,solid},%
{draw=black,mark=+,solid},%
{black,mark=o},}}

\pgfplotsset{tick scale binop=\times}

\usetikzlibrary{shapes,decorations,arrows,calc,fit,positioning,pgfplots.groupplots}
\tikzset{
    auto,node distance =1 cm and 1 cm,semithick,
    state/.style ={ellipse, draw, minimum width = 0.7 cm},
    point/.style = {circle, draw, inner sep=0.04cm,fill,node contents={}},
}

\newcommand{\final}[1]{\textcolor{black}{#1}}

\newtheorem{hypo}{Hypothesis}
\newtheorem{ass}{Assumption}
\newtheorem{resq}{Research Question}

\usepackage{marginnote}

\author{Amir H. Jadidinejad, Craig Macdonald, Iadh Ounis}
\affiliation{%
  \institution{University of Glasgow}
  \country{UK}
}
\email{firstname.lastname@glasgow.ac.uk}

\setcopyright{acmcopyright}

\begin{document}

\title{The Simpson's Paradox in the Offline Evaluation of Recommendation Systems}

\begin{abstract}

Recommendation systems are often evaluated based on user's interactions that were collected from an existing, already deployed recommendation system. 
In this situation, users only provide feedback on the exposed items and they may not leave feedback on \final{other} items \final{since} they have not been exposed to them by the deployed \final{system}. 
As a result, the collected feedback dataset \final{that} is used to evaluate a new model is influenced by the deployed \final{system}, \final{as a form of} closed loop feedback.
In this paper, we show that \final{\final{the} typical offline evaluation} \final{of recommender systems} suffers from \final{the so-called} Simpson's \final{paradox}. Simpson's \final{\final{paradox} is the name given to a phenomenon observed when} a significant 
trend appears in several different sub-populations of observational data but disappears or \final{is} even \final{reversed} when these sub-populations are combined together.
Our in-depth experiments based on stratified sampling reveal that a \final{very small} \final{minority of items that are \final{frequently} exposed} by the deployed \final{system} plays a confounding factor in the offline evaluation of recommendation systems.
In addition, we propose a novel evaluation methodology that takes into account the confounder, i.e\ the deployed \final{system's} \final{characteristics}.
\final{Using the relative comparison of many recommendation models as in the typical offline evaluation of recommender systems, \final{and based on \final{\final{the} Kendall rank correlation coefficient}}, we show that our proposed evaluation \final{methodology} \final{exhibits} \final{\final{statistically} significant improvement\final{s} of} 14\% \final{and 40\% on \final{the} \final{examined open loop datasets (Yahoo! and Coat), respectively},}  in reflecting the true ranking of systems with an open loop (randomised) evaluation in comparison to the standard evaluation}.

\end{abstract}



\maketitle

\section{Introduction}
\label{sec:Introduction}

Recommendation systems \final{are typically} measured in either \final{an} {\em online} (A/B testing or interleaving) or \final{an} {\em offline} manner. However, the offline evaluation of recommender systems remains the most widely reported evaluation in the literature \final{due to the difficulties in deploying research systems for online evaluation}~\cite{recsys_offline_options20,evaluating_recsys04}. In reality, it is hard to \final{reliably} evaluate the effectiveness of a recommendation model in an offline fashion using logs of historical user interactions~\cite{recsys_eval19, algorithmic_confounding18,ir_metrics_robustness_recsys18}. The issue is that there may be {\em confounders}, i.e.\ variables that affect both the {\em exposure} of items and the {\em outcomes} (e.g.\ ratings)~\cite{deconfound_recsys18}. Confounding \final{problems occur} when a platform attempts to model user behaviour without taking into account the \final{selection bias in the exposed recommended items~\cite{expected_exposure20}}, leading to unintended consequences. In this situation, it is difficult to distinguish which user interactions stem from \final{the} users' true preferences and which are influenced by the \final{deployed recommender system}.

Typically, the offline evaluation of recommendation systems has two stages: (1) gathering a collection of user's feedback from a deployed system; and (2) the use of such feedback to empirically evaluate and compare different recommendation models. 
The first stage can be subject to different types of confounders, either initiated by the item selection behaviour of users or through the actions of the deployed recommendation system~\cite{closed_loop_feedback_sigir20,algorithmic_confounding18,offline_eval_spotify19}. 
For instance, \final{aside from other interaction mechanisms such as search, users are more likely to interact with those items that they have been exposed to.}
In this way, \final{using historical interactions in the offline evaluation of a new recommender model, where those interactions have been \final{obtained} from the deployed recommender system}, \final{forms} \final{a {\em closed \final{loop} feedback}}~\cite{closed_loop_feedback_sigir20,debiasing_feedback_loop19}, i.e.\ the deployed recommender system\footnote{\final{Naturally, a recommender system relies on a recommendation model or algorithm. Hence, in this paper, we interchangeably use system and model.}} has a direct \final{effect} on the collected feedback, which is \final{in turn} used for the offline evaluation of other recommendation models.
\final{\final{Hence, the} new recommendation model is \final{evaluated in terms of how much it tends to mimic the interactions collected by} the deployed model, rather than \final{how well it meets} \final{the true preferences of users}. On the other hand, in an {\em open loop (randomised)} scenario, the deployed model is a {\em random} recommendation model, i.e.\ random items are exposed to the \final{users}. Therefore, the feedback loop between the deployed model and the new recommendation model is broken and the deployed model has no effect on the collected feedback dataset \final{and, consequently, it has also no effect on} the offline evaluation of any new model based on the collected feedback dataset.}
\final{However, it \final{is naturally not practical} to expose users to random items \final{due to the potential for degraded user experience}. %
Therefore, \final{the training and evaluation} of recommendation systems based on closed loop feedback is a critical problem. }

Simpson's paradox is a phenomenon in statistics, in which a significant trend appears in several different groups of observational data but disappears or even reverses when these groups are combined together~\cite{understanding_simpson_paradox14}. \final{In recommendation scenarios,}
\final{the} Simpson's paradox occurs when the association between the exposure (i.e.\ the recommended items) and the outcome (i.e.\ \final{the} user's explicit or implicit feedback) is investigated but the exposure and outcome are strongly \final{influenced \final{by}} a third variable. \final{In statistics,} \final{if the observed feedback \final{is} a product of Simpson's paradox, \final{the stratification of} the feedback according to the confounding variable \final{causes the disappearance of} the paradox.} %
\final{We argue that} in the case of recommendation systems, this confounding variable is the deployed \final{model} (\final{or system}) from which the interaction data were collected, a.k.a.\ closed loop feedback~\cite{closed_loop_feedback_sigir20}. 
\looseness -1 \final{Our key aims in this paper are to provide an in-depth investigation of the} impact of closed loop feedback on the offline evaluation of recommendation systems \final{and to provide a robust solution to the problem}. \final{In particular, we} argue that the feedback datasets collected from a deployed model are biased towards the deployed model's \final{characteristics}, leading to the conclusions being prone to \final{the} Simpson's paradox. We \final{thoroughly} investigate the effect of the confounding variable (i.e.\ the deployed model's \final{characteristics}) on the offline evaluation of recommendation systems \final{where significant trends are \final{observed}, which} \final{then} \final{disappear} or even \final{reverse} \final{when reporting observations from the classical offline setting}. \final{In addition, }\final{we propose a new evaluation methodology that addresses the Simpson's paradox in order to conduct a sounder offline evaluation of recommendation systems.}

To better understand the subtleties of \final{the tackled} problem, consider a deployed recommendation model that promotes a specific group of items (e.g.\ popular items\footnote{It is well-known in the literature that recommendation algorithms and public datasets favor popular items~\cite{recsys_offline_options20,popularity_accuracy11,recsys_eval19}.}) -- there are a few head items that are widely exposed to the users \final{that} obtain \final{a large number} of interactions, in contrast to a long tail of items with \final{only a few} interactions. 
When we evaluate a new recommendation model based \final{on} feedback that \final{was} collected from the \final{afore}mentioned deployed model without taking into account \final{(or adjusting)} {\em how \final{frequently} different items \final{were} exposed}, the evaluation process is confounded by the deployed model's \final{characteristics}, i.e.\ the performance of any model that has a tendency to \final{exhibit similar properties to the already deployed model} \final{will likely} be over-estimated.
\final{In this situation, we might choose to deploy a new model that matches the properties of the already deployed model, while it might be less effective than another model \final{from the \final{actual} users' perspective}.}
\final{In this paper, we thoroughly investigate the consequences of this problem on the conclusions drawn from the \final{standard} offline evaluation of recommender systems and propose a new methodology to address the issue}.  \final{In particular,} \final{the \final{contributions} of this paper \final{are} \final{two-fold}}:
\begin{itemize}
    \item We present an in-depth analysis of the \final{standard} offline evaluation of recommender systems, showing that it \final{indeed} suffers from \final{the} Simpson's paradox. 
    Our in-depth experiments on \final{four} well-known closed loop and open loop (randomised) datasets reveal that a {\em \final{very small} minority} of items that are over-exposed by the deployed model play a confounding factor in \final{a standard offline evaluation}, i.e.\ significant trends in a \final{given} recommender system's performance that \final{are} observed in the {\em majority} of user-item interactions actually disappear or even reverse in the \final{standard} offline evaluation when we merge this majority feedback with a minority confounder. 
    \item \looseness -1 To address this problem, we propose a \final{new} propensity-based stratified evaluation method in comparison to \final{both} the \final{standard offline} holdout evaluation~\cite{holdout_evaluation10} and \final{the recently introduced} counterfactual evaluation~\cite{rec_treatment16,unbiased_offline_rec_eval18} methods. Our experiments on a broad range of recommendation models and evaluation metrics reveal that the proposed propensity-based stratified evaluation method better represents the actual utility of the examined recommendation models. 
\end{itemize}

The remainder of the paper is organised as follows: In Section~\ref{sec:Related Work}, we position our contributions in comparison to the \final{existing} literature \final{while Section~\ref{sec: Offline Evaluation Methods} describes the current offline evaluation methods. }
Section~\ref{sec:Simpson's Paradox in Recommender Systems} \final{presents} the Simpson's paradox \final{and how it affects the offline evaluation of recommender systems}; \final{Section~\ref{sec:Propensity-based Stratified Evaluation} \final{describes} the \final{new} proposed evaluation method; Section~\ref{sec:Experimental Setup} and Section~\ref{sec:Experimental Results and Analysis} \final{present} the experimental setup \final{as well as} the conducted experiments and their analysis, respectively}. Finally, Section~\ref{sec:Conclusions} \final{summarises} our findings.

\section{Related Work}
\label{sec:Related Work}

\final{Our work is inspired by three lines of related research, ranging from tackling bias in learning to rank methods (Section~\ref{sec: Unbiased Learning to Rank}), through algorithmic confounding or closed loop feedback (Section~\ref{sec: Algorithmic Confounding}), to recent work on counterfactual learning and evaluation (Section~\ref{sec: Counterfactual Learning and Evaluation})}.

\subsection{Unbiased Learning to Rank}
\label{sec: Unbiased Learning to Rank}

Users' feedback in recommendation systems is reminiscent of clickthrough data in information retrieval (IR). Previous research~\cite{ir_evaluation_clickthrough11} has shown that there is a strong dependency between the documents\footnote{These correspond to items in recommendation systems.} presented to the user, and those for which the system receives feedback, i.e.\ higher-ranked documents obtain more clicks (\final{a phenomenon denoted in the literature as} position/presentation bias). 
For example, consider a movie recommendation system that aims to learn the users' preferences by observing their watching history. The system can infer which \final{movie(s)} they prefer over other programs at a particular time, but these are \textit{relative preferences} in relation to the exposed movies, \final{but} \final{do not constitute} preferences on an \textit{absolute} scale.
Previous research \final{in information retrieval (IR)}~\cite{clickthrough_retrieval_quality08,ir_evaluation_clickthrough11} has revealed that such feedback does not reliably reflect retrieval quality while \final{nonetheless still} providing relative \final{user} preferences in paired experiments. 
Classical \final{offline evaluation \final{setups} in IR} use {\em pooling} \final{to gather a comprehensive set of possibly relevant documents for assessment, thereby preventing bias in the relevance assessments} towards a single system (known as incompleteness). 
However, pooling is not a suitable approach for addressing incompleteness in the evaluation of recommendation systems, due to the contextual and personalised nature of \final{users'} recommendation needs.

\looseness -1 \final{\final{Recent,} unbiased learning to rank approaches aim to learn reliable relevance signals from noisy and biased clickthrough data either with the help of click models or \final{through} counterfactual methods~\cite{propensity_svm_rank17,unbiased_ltr_theory_practice18,unbiased_ltr_ubiased_propensity18,policy_aware_unbiased_ltr20}. 
These models typically require a large quantity of clicks, \final{which} makes them difficult to apply in systems where click data is highly sparse due to \final{the} personalised nature of recommendation systems~\cite{l2r_selection_bias16,selection_bias_ltr20}.
On the other hand, these models are designed to tackle position/presentation bias in IR systems. \final{However, 
although} \final{both} position and presentation bias can influence the collected feedback from a deployed recommendation system, recommendation systems mostly suffer from selection bias~\cite{closed_loop_feedback_sigir20,algorithmic_confounding18,degenerate_feedback_loop19}.}
\final{In this paper, we investigate the origin of selection bias in recommendation systems and provide a novel evaluation method that takes into account the role of the deployed recommendation system.}

\subsection{Algorithmic Confounding}
\label{sec: Algorithmic Confounding}

Recommendation systems are often evaluated with \final{interaction} data \final{obtained} from users that have already been exposed to a deployed algorithm. This creates a \textit{feedback loop}~\cite{degenerate_feedback_loop19,debiasing_feedback_loop19} (a.k.a.\ \textit{algorithmic confounding}~\cite{algorithmic_confounding18} or closed loop feedback~\cite{closed_loop_feedback_sigir20}). 
Using simulations on synthetic data, Chaney et al.~\cite{algorithmic_confounding18} revealed that this feedback loop in recommendation models \final{encourages} similar users to interact with the same set of items, \final{thereby} homogenising their behavior, relative to \final{an open loop (randomised)} platform \final{where random items are exposed to the user}.
In contrast to the simulation experiments, Jadidinejad et al.~\cite{closed_loop_feedback_sigir20} investigated the effect of closed loop feedback in both the training and offline evaluation of recommendation models based on \final{an} \final{open loop (randomised)} dataset. They found that there is a strong correlation between the deployed model \final{from which} the \final{users'} feedback were collected and a new model that is trained or evaluated based on the collected feedback dataset. For example, when popular items are favoured by the deployed algorithm, the observed interactions are biased towards popular items. In this situation, the performance of any algorithm that has a tendency to recommend popular items might be over-estimated~\cite{algorithmic_confounding18,recsys_eval19,closed_loop_feedback_sigir20}. 
\final{In this paper, we propose a theoretical framing of feedback loops in recommendation systems based on the Simpson's paradox, \final{as well as formulate and validate} hypotheses about the \final{implications} of closed loop feedback in the offline evaluation of recommendation systems. In addition, we \final{propose} a novel evaluation methodology, \final{which takes} into account the confounding role of the deployed model \final{and show its benefits in comparison} to both the \final{standard offline} holdout evaluation~\cite{holdout_evaluation10} and the recently introduced counterfactual evaluation~\cite{rec_treatment16,unbiased_offline_rec_eval18} methods}.

\subsection{Counterfactual Learning and Evaluation}
\label{sec: Counterfactual Learning and Evaluation}

Although researchers in IR have proposed novel approaches for both learning~\cite{propensity_svm_rank17} and evaluating~\cite{clickthrough_retrieval_quality08,ir_evaluation_clickthrough11} from user's clickthrough data, many researchers and practitioners \final{still} focus on evaluating recommendation systems in terms of accuracy metrics on datasets that were collected from an already deployed system~\cite{recsys_offline_options20,evaluating_recsys04,recsys_missing_not_random10,popularity_accuracy11}. 
On the other hand, \final{several recent studies}~\cite{counterfactual_eval_learn16,rec_treatment16,unbiased_offline_rec_eval18} \final{have} attempted to estimate the actual utility of a recommendation model with propensity-weighting techniques commonly used in the \textit{off-policy evaluation} approaches in Reinforcement Learning with the aim of achieving an unbiased performance estimator of biased observations.
\final{Propensity in this context is the probability that the deployed model \final{has exposed} item $i$ to user $u$ by the time of feedback collection.}
The intuition \final{in counterfactual evaluation} is that each feedback is \final{individually} weighted based on its propensity. 
The key to a fair evaluation in \final{the} counterfactual approaches lies in accurately logging propensities in \final{an {\em open loop (randomised)} dataset where random items are exposed to the user}. \final{However, most current public recommendation datasets do not introduce a randomisation element}~\cite{rec_treatment16,unbiased_offline_rec_eval18}, \final{making it practically difficult to effectively deploy the counterfactual approaches for evaluating recommender systems}.
\final{In the absence of open loop (randomised) feedback, the estimated propensities are noisy and skewed. As a result, weighting each feedback individually based on its propensity score \final{can lead} to \final{a} high variance \final{in the estimation of \final{the} model's effectiveness}, as \final{reported} in \final{several} previous \final{studies}~\cite{ batch_logged_bandit_counterfactual15,counterfactual_eval_learn16,rec_treatment16,unbiased_offline_rec_eval18}}.
\final{Hence, compared} to previous \final{works} on propensity weighting~\cite{rec_treatment16,unbiased_offline_rec_eval18}, our model does not weight each feedback individually based on its propensity. We \final{instead} propose a novel evaluation method based on \textit{stratification}, \final{which} \final{\final{considers} each feedback in the corresponding stratum based on the estimated propensities.}
As a result, \final{as our experiments show,} \final{our} proposed evaluation method is more robust \final{in} \final{\final{coping with these} noisy \final{estimated}} propensities.

\section{Offline Evaluation Methods}
\label{sec: Offline Evaluation Methods}

\final{In this section, we summarise the current offline evaluation methods commonly used in the literature, namely \final{the \final{standard}} holdout evaluation (Section~\ref{sec:Holdout evaluation}) and \final{the} counterfactual evaluation (Section~\ref{sec:Counterfactual Evaluation}).}

\subsection{Holdout Evaluation}
\label{sec:Holdout evaluation}

The purpose of \final{the} holdout evaluation~\cite{holdout_evaluation10} is to test a model on \final{a} different data \final{from what it} was trained on. In this method, the dataset is {\em randomly} divided into \final{a training set and a test set, the latter being the holdout} data. The effectiveness of a model ($\hat{R}$) for the predicted item ranking $\hat{Z}$ can be calculated as follows on the holdout data~\cite{unbiased_offline_rec_eval18}:
\begin{equation}
\hat{R}_{\mathrm{holdout}}(\hat{Z})=\frac{1}{|\mathcal{U}|} \sum_{u \in \mathcal{U}} \frac{1}{\left|\mathcal{S}_{u}^{*}\right|} \sum_{i \in \mathcal{S}_{u}^{*}} c(\hat{Z}_{u, i})
\end{equation}
where $\mathcal{S}_{u}^{*}$ is the set of items (among all the items in $\mathcal{I}$) \final{that are} exposed to the user $u \in \mathcal{U}$, \final{$\hat{Z}_{u, i}$} is the predicted ranking of the exposed item $i$ for user $u$, and the function $c$ denotes any top-k scoring metric, such as normalised Discounted Cumulative Gain \final{at rank cutoff k} (nDCG\final{@k}).

The holdout evaluation method is the \final{standard and} most common evaluation method in the field of recommendation systems because of its simplicity and flexibility~\cite{recsys_offline_options20,evaluating_recsys04}. However, if the dataset \final{is} biased towards a group of items, the holdout test set will be a {\em biased sample} of the user-item interactions~\cite{deconfound_recsys18}. Therefore, the holdout evaluation does not necessarily represent the actual effectiveness of the examined models, i.e.\ the expected outcome of the holdout evaluation does not \final{necessarily} conform to the open loop evaluation~\cite{unbiased_offline_rec_eval18}:
\begin{equation}
\mathbb{E}_{O}\left[\hat{R}_{\mathrm{holdout}}(\hat{Z})\right] \neq R(\hat{Z})    
\end{equation}
\looseness -1 where $R$ is the actual performance of the predicted \final{ranking} \final{of items} $\hat{Z}$. In this situation, the effectiveness of any model that captures the bias in the original dataset might be over-estimated~\cite{algorithmic_confounding18,unbiased_offline_rec_eval18}. For example, when popular items are favoured by the deployed algorithm, the observed interactions are biased towards popular items. \final{As a result}, the performance of any algorithm that has a tendency to recommend popular items might be over-estimated~\cite{recsys_eval19}.
An expensive solution is to create an open loop (randomised) test set from \final{the} experiments, i.e.\ \final{to} ask \final{the} users to provide feedback for randomly selected items. \final{Another alternative and more practical solution consists in using} \final{the} \final{propensity-based} methods~\cite{propensity_score_methods_survay11} (e.g.\ counterfactual evaluation)\final{\final{, which we describe} in the following section}.

\subsection{Counterfactual Evaluation}
\label{sec:Counterfactual Evaluation}

The problem of evaluating recommendation models using historical feedback is similar to counterfactual evaluation in reinforcement learning. In this setting, the goal is to evaluate a new recommendation model based on a feedback dataset that \final{is} collected from a deployed recommendation model. 
Inverse Propensity Scoring (IPS) leverages importance sampling to account for the fact that the feedback, \final{which was} collected from the deployed recommendation model \final{is} not \final{uniformly} random. \final{Theoretically, the IPS evaluation method} is able to evaluate a new model \final{independently from} the deployed model (\final{the} confounder). In the context of recommendation systems, \final{the} IPS evaluation method \final{was} proposed for both explicit~\cite{rec_treatment16} and implicit~\cite{unbiased_offline_rec_eval18} feedback. The key idea \final{of} IPS is to reweight feedback based on \final{the} propensities of the logged items using importance sampling. \final{The propensity} $p_{u,i}$ in this context is the probability \final{that} item $i$ \final{would} \final{have} \final{been} shown to the user $u$ \final{by the time \final{the} feedback data \final{was} collected}. 
The effectiveness of a model ($\hat{R}$) for the predicted item ranking $\hat{Z}$ and the provided propensity scores $P$ can be calculated as follows~\cite{unbiased_offline_rec_eval18}:
\begin{equation}
\hat{R}_{\mathrm{IPS}}(\hat{Z} | P)=\frac{1}{|\mathcal{U}|} \sum_{u \in \mathcal{U}} \frac{1}{\left|\mathcal{S}_{u}^{\final{*}}\right|} \sum_{i \in \mathcal{S}_{u}^{*}} \frac{c(\hat{Z}_{u, i})}{p_{u, i}}
\end{equation}

The intuition behind \final{the} IPS method is that the propensity score $p_{u,i}$ is a balancing factor: \final{conditioned} on the propensity score, the \final{observed feedback} distribution will be \final{adjusted to} \final{the examined model.}
Unlike \final{the} holdout evaluation $\hat{R}_{\mathrm{holdout}}$, the IPS evaluation method is theoretically unbiased ($\mathbb{E}_{O}\left[\hat{R}_{\mathrm{IPS}}(\hat{Z})\right] = R(\hat{Z})$) if and only if we can sample open loop feedback from all possible items $\mathcal{I}$ {\em uniformly at random}~\cite{batch_logged_bandit_counterfactual15}. However, in various situations, it may be difficult or even impossible to expose users to random items due to \final{obvious} business constraints \final{(users are unlikely to tolerate that the system keeps showing them random non-relevant items)}. \final{Moreover}, the imbalance of the assigned propensities will allow for some feedback to be weighted disproportionately. As a result, a major difficulty in applying \final{the} IPS evaluation method in practice is its high variance \final{in the estimation of \final{a given} model's effectiveness}~\cite{batch_logged_bandit_counterfactual15}. 
In the following, we investigate the impact of \final{our} proposed propensity-based stratified evaluation in comparison to both \final{the} holdout and counterfactual evaluation.

\section{Simpson's Paradox}
\label{sec:Simpson's Paradox in Recommender Systems}

Simpson's paradox \final{is the name given to an observed} phenomenon in statistics, in which a significant trend appears in several different groups of observational data but disappears or even reverses when these groups are combined together.
This topic has been widely studied in the causal inference \final{literature}~\cite{understanding_simpson_paradox14,simpson_psy13}. 
In this phenomenon, an apparent paradox arises because aggregated data can support a conclusion \final{that} is opposite \final{to the conclusions} from the same stratified data before aggregation.
\final{\final{The} Simpson's paradox occurs when the association between two variables is investigated but these variables are strongly \final{influenced} \final{by} a confounding variable. \final{When the data is stratified} according to the confounding variable, \final{the paradox is revealed showing paradoxical conclusions}.}
\final{In this situation, \final{the}  use of a significance test \final{might} identify unsound conclusions made across a specific stratum; however, as we show later in Section~\ref{sec:Experimental Results and Analysis}, significance tests may not \final{possibly} identify such trends. Indeed, in a recommender system evaluation scenario, testing occurs over users, while the paradox we discuss here usually \final{involves} the user-item feedback generation process.} \final{On the other hand}, the Simpson's paradox can be resolved when causal relations are appropriately addressed \final{in} statistical modeling~\cite{understanding_simpson_paradox14}.

\begin{table}
\centering
\caption{\final{Simpson's paradox examples in medical studies (a) and recommendation systems (b).}}
    \begin{subtable}{.45\textwidth}
        \centering
        \caption{Kidney stone example~\cite{kidney_stone86}. Each entry represents the number of responses out of the total number of patients \final{while \final{the} success rates are shown in brackets. \newline \newline }}
        \label{tbl:simpson_example}
        \begin{tabular}{l|c|c}
            & Treatment A & Treatment B \\ \hline
            \multirow{2}{*}{Size $<$ 2cm} & Group 1 & Group 2 \\ 
            &  \phantom{0}81/87\phantom{0} (\textbf{93\%}) &  234/\textbf{270} (87\%) \\ \hline
            \multirow{2}{*}{Size $\geq$ 2cm} & Group 3 & Group 4 \\ 
            &  192/\textbf{263} (\textbf{73\%}) &  \phantom{0}55/80\phantom{0} (69\%) \\ \hline
            Both & 273/350 (78\%) & 289/350 (\textbf{83\%}) \\ 
        \end{tabular}
    \end{subtable}\hspace{10mm}%
    \final{\begin{subtable}{.45\textwidth}
    \centering
    \caption{\final{Simpson's paradox in the offline evaluation of recommender systems. Each entry represents the effectiveness of the examined model on the corresponding stratum.} \final{* denotes a significant difference compared to the other model (paired t-test, p <0.05).}}
    \label{tbl:simpson_recommender}
    \begin{tabular}{l|c|c}
        & \phantom{***}Model A\phantom{***} & \phantom{***}Model B\phantom{***} \\ \hline
        \multirow{2}{*}{Q1 (\textbf{99\%})} & Group 1 & Group 2 \\
        & 0.339 & \textbf{0.350}* \\ \hline
        \multirow{2}{*}{Q2 (\textbf{1\%})} & Group 3 & Group 4 \\
        & \textbf{0.695}* & 0.418 \\ \hline
        Both & \textbf{0.373}* & 0.369 \\ 
    \end{tabular}
    \end{subtable}}
\end{table}

In this section, \final{to illustrate the Simpson's paradox}, we present a real-life example from a seminal paper~\cite{kidney_stone86} comparing the success rates of two treatments for the kidney stone disease.\footnote{Treatments A and B \final{correspond} to `All open procedures' and 
`Percutaneous nephrolithotomy' in the original paper~\cite{kidney_stone86}.} 
The objective is to find out which treatment is more effective based on observational studies. 
Charig et al.~\cite{kidney_stone86} randomly sampled 350 patients who were exposed to each treatment and reported the success rates in Table~\ref{tbl:simpson_example}. 
A plausible conclusion is that treatment B is more effective than treatment A (83\% vs. 78\% recovery rate). On the other hand, the paradoxical conclusion is that the success rates reverse when \final{the} stone size is taken into account, i.e.\ treatment A is more effective for both small (93\% vs. 87\%) and large stone sizes (73\% vs. 69\%). 
Charig et al.~\cite{kidney_stone86} argued that the treatment (A vs. B) and the outcome (success vs. failure) are strongly associated with a third confounding variable (\final{namely, the} stone size):

\begin{hypo}
\final{Doctors} \final{tend to opt for} favoring treatment B for less severe cases (i.e.\ small stones) while \final{favouring} treatment A for more severe cases (i.e.\ large stones).
\label{hyp:simpson_example}
\end{hypo}

Table~\ref{tbl:simpson_example} \final{(data from~\cite{kidney_stone86})} \final{validates} the above hypothesis, i.e.\ the majority of patients who were exposed to treatment A had large stones (263 out of 350 random patients in Group 3) while the majority of patients who were exposed to treatment B had small stones (270 out of 350 random patients in Group 2). As a result, \final{when the samples are randomly selected from the} patients who were exposed to \final{treatments} A or B, \textit{\final{the} sampling process is not \final{purely} random}, i.e.\ \final{the samples are systematically skewed towards the} patients who have severe cases for measuring treatment A, while sampling mild cases for measuring treatment B. \final{This} is a well-known phenomena in causal analysis, which is \final{also} known as the Simpson's paradox~\cite{understanding_simpson_paradox14}.

\final{Table~\ref{tbl:simpson_recommender} \final{shows} \final{an illustrative} example of the Simpson's paradox in the offline evaluation of recommendation systems.}\footnote{\final{\final{This example is actually a highlight of our later experiments}, \final{detailed} further in Section~\ref{sec:Experimental Results and Analysis}.}}
\final{We are interested to evaluate the effectiveness of two recommendation models, depicted as A and B in Table~\ref{tbl:simpson_recommender}. Both models are evaluated on the same dataset and \final{using} the same evaluation metric.} %
\final{The standard offline evaluation on the examined dataset shows that model A \final{significantly} outperforms model B, \final{according to a paired t-test}. However, stratifying the examined dataset into two strata (Q1 and Q2) reveals that indeed model B is the \final{significantly} better model for 99\% of the examined dataset while model A is \final{\final{statistically} superior} only for a \final{small} minority (1\%) of the examined dataset. The \final{statistical} superiority of model \final{B} in 99\% of the examined dataset disappears when aggregating \final{the} Q1 and Q2 \final{strata} together.}
\final{In the following section, we present how \final{the} Simpson's paradox affects the offline evaluation of recommendation systems, discuss the causes \final{of such a paradox} and propose a new evaluation methodology to address \final{it}.}

\section{Propensity-based Stratified Evaluation}
\label{sec:Propensity-based Stratified Evaluation}

\tikzstyle{block} = [draw, fill=blue!20, rectangle, 
    minimum height=2.5em, minimum width=6em]
\tikzstyle{output} = [coordinate]
\tikzstyle{pinstyle} = [pin edge={to-,thin,black}]

\begin{figure}
    \centering
    \begin{subfigure}[b]{0.45\textwidth}
        \resizebox{\textwidth}{!}{%
            \begin{tikzpicture}[auto, node distance=2cm]
                \node [name=input] (input) {};
                \node [block, right of=input] (recsys) {RecSys};
                \node [block, right of=recsys, pin={[pinstyle]above:\final{Other Confounding Variables}},
                        node distance=3cm] (user) {User};
        
                \draw [->] (recsys) -- node[name=u] {$e$} (user);
                \node [output, right of=user] (output) {};
        
                \draw [->] (input) -- node {$r$} (recsys);
                \draw [->] (user) -- node [name=r] {$r$}(output);
                \draw [-] (r) -- ++ (0,-2.1) node [xshift=-85,yshift=10] {Closed Loop Feedback} -|  ++ (-6.40,1.89);
            \end{tikzpicture}
        }%
        \caption{\final{Data flow in a recommender system}~\cite{closed_loop_feedback_sigir20}.}
        \label{fig:closed_loop}
    \end{subfigure}\hfill
    \begin{subfigure}[b]{0.45\textwidth}
        \resizebox{\textwidth}{!}{%
            \begin{tikzpicture}
                \node (1) at (0,0) [align=center] {RecSys' \final{Characteristics} \\ ($e$)};
                \node (2) [below of =1, left of =1, align=center, xshift=-3em, yshift=-2em] {Closed Loop Feedback \\ ($r$)};
                \node (3) [below of =1, right of =1, align=center, xshift=+3em, yshift=-2em] {Offline Evaluation \\ ($Y$)};
            
                \path[->] (2) edge  (3);
                \path[->] (1) edge  (2);
                \path[->,dashed] (1) edge  (3);
            \end{tikzpicture}
        }%
        \caption{Assumed coarse-grained causal diagram.}
        \label{fig:causal}
    \end{subfigure}
    \caption{Closed loop feedback in recommendation systems.}
    \label{fig:closed_loop_all}
\end{figure}
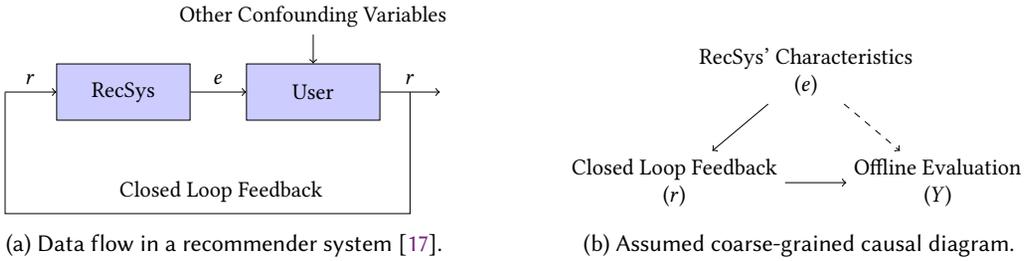

\final{When creating a dataset for the offline evaluation of a recommender system, the users' feedback can be collected not only \final{from interactions with items suggested by} the deployed recommendation system but also through other modalities such as \final{interactions occurring when} browsing the items' catalogue or when following links to sponsored items. It is not straightforward to distinguish the source of the users' feedback \final{since no public datasets provide sufficient data to establish the sources of the users' feedback}. Hence, in this paper we focus our investigation on the main source of user's feedback, namely the deployed system~\cite{algorithmic_confounding18, degenerate_feedback_loop19, recsys_eval19,spotify_diversity20}. } In order to \final{demonstrate} the Simpson's paradox in recommendation systems, we need a causal hypothesis similar to Hypothesis~\ref{hyp:simpson_example} in Section~\ref{sec:Simpson's Paradox in Recommender Systems}.

Figure~\ref{fig:closed_loop} shows the \final{information flow of a} \final{typical} recommendation system \final{where the users' feedback is collected through the deployed system}.
The deployed recommender system \final{component (denoted \final{by} RecSys)} filters items for the target user (e.g. by suggesting a ranked list of items) depicted as \final{the} exposure ($e$). 
On the other hand, the user's recorded preferences \final{(e.g.\ ratings \final{or clicks})} towards items (depicted as $r$) are leveraged as interaction data to {\em train} \final{and/}or {\em evaluate} the next-generation of the recommendation model. \final{Hence, \final{since} the \final{users'} clicks are obtained on items exposed by the RecSys\footnote{\final{Their clicks may also be affected by other confounding variables, such as temporal factors and other browsing interactions with the item catalogue, as illustrated in Figure~\ref{fig:closed_loop}.}},} the model itself influences the generation of data that is used to \final{either} train or evaluate it~\cite{degenerate_feedback_loop19}.
The system presented in Figure~\ref{fig:closed_loop} is a dynamic system where a simple associative reasoning about the system is difficult because each component influences the other. Figure~\ref{fig:causal} shows the corresponding causal diagram in such a closed loop feedback scenario. \final{The solid lines represent an explicit/observed relationship between \final{the} causes and effects while the dashed line represents an implicit/unobserved relationship.}
\final{\final{As mentioned above,} in the case of recommendation systems, \final{the} \final{main} confounding variable is the deployed \final{model} from which the interaction data were collected}~\cite{algorithmic_confounding18, degenerate_feedback_loop19, recsys_eval19}.
Our aim is to \final{evaluate} the effectiveness of a recommendation model ($Y$) based on closed loop feedback ($r$) collected from a deployed model while \final{the} analysis is influenced by the \final{main} confounder, i.e.\ the deployed model's \final{characteristics}. In this situation, it is difficult to \final{distinguish} \final{those} \final{users'} interactions \final{that} stem from \final{the} users' true preferences \final{from those that} are influenced by the deployed recommendation model. \final{As a result,} \final{in the scenario where the users' feedback is typically collected through the deployed system} \final{we postulate that} the offline evaluation of recommendation models \final{based on closed loop feedback datasets} is strongly influenced by the \final{underlying} deployed recommendation model:

\begin{hypo}
Closed loop feedback \final{collected} from a deployed \final{recommendation} model \final{is} skewed towards the deployed model's \final{characteristics} (i.e.\ \final{the} exposed items) and the deployed model's \final{characteristics} \final{play} a confounding factor in the offline evaluation of recommendation models.
\label{hyp:closed_loop}
\end{hypo}

The core problem is that the deployed recommendation model is attempting to model the underlying user preferences, making it difficult to make claims about user behaviour (or use the behaviour data) without accounting for algorithmic confounding~\cite{algorithmic_confounding18, degenerate_feedback_loop19}.
On the other hand, if the RecSys component in Figure~\ref{fig:closed_loop} is a {\em random} model, then the interconnection between the User and the RecSys components \final{in Figure~\ref{fig:closed_loop}} will be removed and the RecSys component has no effect on the collected feedback dataset, \final{a situation} known as {\em open loop feedback}, i.e.\ no item \final{receives} more or less expected exposure compared to any other item in the collection~\cite{expected_exposure20}.

\final{As shown in Figure~\ref{fig:causal}, if the deployed model's characteristics ($e$) could be identified and measured, the offline evaluation ($Y$) 
would be independent \final{of the confounder}\final{.} %
Therefore, i}n order to \final{validate} the effect of closed loop feedback in a recommendation system (Hypothesis~\ref{hyp:closed_loop}), we have to quantify the deployed model's \final{characteristics}. To this end, \final{following~\cite{propensity_score_methods_survay11,unbiased_offline_rec_eval18},} we define the propensity score as follows:

\begin{definition}
  The propensity score $p_{u,i}$ is the tendency of the deployed model (depicted as RecSys in Figure~\ref{fig:closed_loop}) to expose item $i\in\mathcal{I}$ to \final{user} $u\in\mathcal{U}$.
  \label{def:propensity_score}
\end{definition}

\final{The propensity} $p_{u,i}$ is the probability that item $i$ \final{is} exposed to user $u$ by the deployed model \final{under a closed loop feedback scenario}\footnote{In Section~\ref{sec:Estimating Propensity Scores} we propose a method to estimate the propensity scores in our experiments.}. 
This value quantifies a system's deviation from an \final{unbiased} open loop exposure \final{ scenario}, \final{where random items are exposed to the user and the deployed model has no effect on the collected feedback.}
\final{The propensity score $p_{u,i}$} allows us to design and analyse the offline evaluation of recommendation models  \final{based on the observed} closed loop feedback so that it mimics some of the particular \final{characteristics} of the open loop scenario.

Stratification is a well-known approach to identify and estimate causal effects \final{by \final{first} identifying \final{the} underlying strata \final{before  investigating} causal effects in each stratum}~\cite{propensity_score_methods_survay11}.
The general intuition is to stratify on the confounding variable and \final{to} investigate the potential outcome in each stratum. 
As a result, \final{it becomes possible to} measure the potential outcome irrespective of the confounding variable. 
\final{In this situation,} a \final{marginalisation over the confounding} variable 
can be leveraged as a combined \final{estimate}.
\final{As mentioned before, the hypothetical confounding variable in recommender systems is the deployed model's characteristics (Hypothesis~\ref{hyp:closed_loop}).  \final{Definition~\ref{def:propensity_score} quantifies} this variable as propensities. In this situation, stratification on the propensity scores allows us to analyse the effect of the deployed model's characteristics on the offline evaluation of recommendation models.}

For the sake of simplicity, suppose that we have a single \final{categorical confounding} variable $\mathrm{X}$.
If we stratify the \final{observed} outcome based on the possible values of $\mathrm{X}$, then the expected value of the potential outcome ($\mathrm{Y}$) is defined as \final{follows}:
\begin{equation}
\mathrm{E}(\mathrm{Y})=\sum_{x} \mathrm{E}\left(\mathrm{Y} | \mathrm{X}=x\right) \mathrm{P}(\mathrm{X}=x)
\label{eq:marginalisation}
\end{equation}
where $\mathrm{E}\left(\mathrm{Y} | \mathrm{X}=x\right)$ is the conditional expectation of the observed outcome given  \final{stratum $x$}, and $\mathrm{P}(\mathrm{X}=x)$ is the marginal distribution of $x$.
\final{For example, in the} kidney stone example presented in Table~\ref{tbl:simpson_example}, the \final{confounding} variable is \final{the} kidney stone size \final{based on which} our observations were stratified ($X=\{small, large\}$)\footnote{\final{Stone size is a continuous variable. In this example, we cast it into a categorical variable for a more clear illustration.}} and the potential outcome is the treatment effect. We can calculate the expected value of each treatment based on Equation~\eqref{eq:marginalisation}. For example, the expected value of treatment A can be calculated as follows:
\begin{equation}
\mathrm{E}(\mathrm{A})= \mathrm{E}\left(\mathrm{A} | \mathrm{X}=small\right) \mathrm{P}(\mathrm{X}=small)+\mathrm{E}\left(\mathrm{A} | \mathrm{X}=large\right) \mathrm{P}(\mathrm{X}=large)
\label{eq:marginalisation_kidney_example}
\end{equation}
\final{Based on the numbers} in Table~\ref{tbl:simpson_example} \final{and Equation~\eqref{eq:marginalisation_kidney_example}}, the expected value of treatment A \final{(resp. B) is} \final{calculated to be} \final{0.832} and \final{0.782}, respectively, i.e.\ $\mathrm{E}(\mathrm{A}) > \mathrm{E}(\mathrm{B})$, which better estimates the actual performance of \final{the} treatments. 
\final{In a similar \final{manner}, for the recommendation example in Table~\ref{tbl:simpson_recommender}, the expected \final{values} of model A and model B \final{are} calculated to be 0.343 and 0.351, respectively, i.e.\ $\mathrm{E}(\mathrm{B}) > \mathrm{E}(\mathrm{A})$. 
}
\final{As mentioned \final{above}, \final{in recommender systems}, the \final{main} hypothetical confounding variable \final{is} the deployed model (Hypothesis~\ref{hyp:closed_loop}). \final{This variable is} quantified as propensity scores (Definition~\ref{def:propensity_score}). Propensity is a continuous variable. In this paper, we cast it into a categorical variable by sorting and segmenting the propensity scores into a predefined number of strata, i.e. Q1 and Q2 strata in Table~\ref{tbl:simpson_recommender}}.
\final{In both cases \final{of Tables~\ref{tbl:simpson_example} and ~\ref{tbl:simpson_recommender}}, stratification based on the hypothetical confounding variable and marginalisation on the examined strata \final{using} Equation~\eqref{eq:marginalisation}, \final{does resolve} the Simpson's paradox. For example, in Table~\ref{tbl:simpson_recommender}, model B should be recognised as the superior model by any reasonable \final{evaluation outcome since} it performs better than model A for 99\% of the user-item feedback \final{on} the examined dataset. This important trend \final{is captured by} the proposed stratified evaluation while \final{it is completely reversed} in the standard offline evaluation when combining the Q1 and Q2 \final{strata} together.
In the following section, we investigate the effect of \final{the} Simpson's paradox and the usefulness of \final{our} proposed propensity-based stratified evaluation in the offline evaluation of recommendation systems. Specifically, we investigate the following research questions:}

\begin{resq}
To what extent is the offline evaluation of recommendation systems affected by the deployed model's \final{characteristics} in a closed loop feedback scenario?
\label{rq:1}
\end{resq}

\final{Here,} our \final{objective} is to assess the \final{confounding} effect of the deployed model's \final{characteristics} on the offline evaluation of recommendation models, as depicted in Figure~\ref{fig:causal}. 
In this regard, similar to \final{the} \final{schematic} example\final{s} presented in Section~\ref{sec:Simpson's Paradox in Recommender Systems}, we are interested to stratify \final{the} observed closed loop feedback based on the deployed model's \final{characteristics}. Such a stratified analysis enables us to assess the presence of \final{the} Simpson’s paradox in the standard offline evaluation of recommendation systems. 
\final{As such}, we \final{investigate} \final{significant trends} in the relative performances of \final{many} different recommendation models \final{where} \final{a trend} is observed in the majority of strata, \final{but the} trend disappears or even reverses in the standard offline evaluation.

\begin{resq}
\final{Can the} propensity-based stratified evaluation help \final{in better estimating} the actual model’s performance \final{when conducting a comparative offline evaluation}?
\label{rq:2}
\end{resq}

\looseness -1 \final{Our objective in \final{the above} research question} is to assess the effectiveness of the proposed propensity-based stratified evaluation in the offline evaluation of recommendation models. As \final{shown} in Equation~\eqref{eq:marginalisation}, we can leverage \final{marginalisation} over the distribution of the confounding variable as a combined estimation of the potential outcome. In this research question, we are interested to assess how well this estimation is \final{correlated} with the open loop \final{(randomised)} evaluation \final{where the deployed model is a random recommendation model, i.e.\ random items are exposed to the user.}
\final{Specifically, \final{given a particular evaluation method (open loop, closed loop, etc.)} we \final{measure the effectiveness of} many recommendation models based on standard rank-based \final{evaluation} metrics \final{(e.g.\ nDCG)} \final{and record the relative ranking of those models in terms of the used metric}. \final{We then investigate the correlation between the relative \final{rankings} of the examined models, comparing the rankings obtained from \final{an} open loop (randomised) setup \final{to} those obtained using our} proposed propensity-based evaluation method.} 
\final{We \final{leverage \final{suitable} significance} tests, as \final{described} in Section~\ref{sec:Experimental Setup}, to \final{determine if} the comparative performances of the examined models predicted by the proposed propensity-based stratified evaluation method better correlates with the open loop (randomised) evaluation in comparison to the standard offline holdout evaluation.} 
Answering this research question provides a better understanding about the actual effectiveness of a given recommendation model in a closed loop scenario.

\section{Experimental Setup}
\label{sec:Experimental Setup}

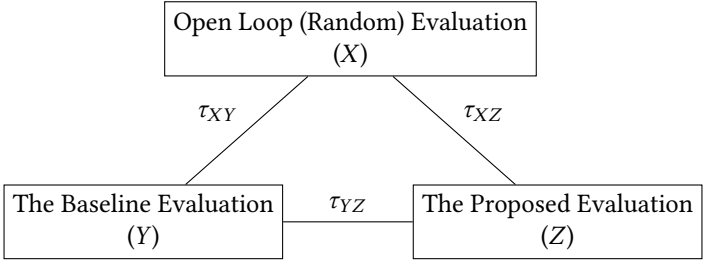
\begin{figure}
    \centering
    \begin{tikzpicture}
        \node (x) at (0,0) [draw, align=center] {Open Loop (Random) Evaluation \\ ($X$)};
        \node (y) [draw, below of =1, left of =1, align=center, xshift=-5em, yshift=-4em] {The Baseline Evaluation \\ ($Y$)};
        \node (z) [draw, below of =1, right of =1, align=center, xshift=+5em, yshift=-4em] {The Proposed Evaluation \\ ($Z$)};
    
        \path (y) edge node {$\tau_{XY}$} (x);
        \path (z) edge node[above right] {$\tau_{XZ}$} (x);
        \path (y) edge node {$\tau_{YZ}$}  (z);
    \end{tikzpicture}
    \caption{The dependent correlations significance test based on Steiger's method~\cite{steiger80}. The effectiveness of each evaluation method ($Y$ and $Z$) is measured based on Kendall's $\tau$ rank correlation with the open loop (\final{randomised}) evaluation ($X$). The significance difference between the baseline evaluation method ($Y$) and the proposed evaluation method ($Z$) is measured based on Steiger's method~\cite{steiger80}.}
    \label{fig:sig_test}
\end{figure}

\looseness -1 \final{In the following, we conduct experiments to address the \final{two} aforementioned research questions.} 
Figure~\ref{fig:sig_test} \final{presents} the general structure of our experiments. 
\final{Each evaluation method ($X$, $Y$ and $Z$) ranks \final{various} examined \final{recommendation} models based on their \final{relative}  \final{performances}.}
We use Kendall's $\tau$ rank correlation coefficient to measure the \final{correlation between the relative order of the examined models in each }evaluation \final{method} ($Y$ or $Z$) in comparison to the ground truth, i.e.\ open loop (\final{randomised}) evaluation ($X$). These correlation values are depicted as $\tau_{XY}$ and $\tau_{XZ}$ in Figure~\ref{fig:sig_test}. 
In addition, we use Steiger's method~\cite{steiger80} to \final{test the difference} between two evaluation methods \final{for significance}, e.g.\ the baseline evaluation method ($Y$) and the proposed evaluation method ($Z$). We compare the effectiveness of \final{our} proposed propensity-based stratified evaluation with both \final{the standard offline} holdout and \final{the} counterfactual evaluation \final{methods} as \final{baselines}.
\final{In \final{the following,} we describe \final{the details of} our experimental setup including \final{the used} datasets and evaluation metrics (Section~\ref{sec:Datasets and Evaluation metrics}), the examined recommendation models (Section~\ref{sec:Recommendation Models}) and \final{how we} estimate \final{the} propensities (Section~\ref{sec:Estimating Propensity Scores}) in our experiments. }

\subsection{Datasets and Evaluation \final{Metrics}}
\label{sec:Datasets and Evaluation metrics}

We use \final{four} datasets, which are widely used in the offline evaluation of recommendation system, namely \final{the} MovieLens~\cite{movielens15}, \final{Netflix~\cite{mf_netflix09}}, Yahoo! music rating~\cite{collaborative_ranking_non_random_missing09} \final{and Coat shopping~\cite{rec_treatment16} datasets}.
MovieLens contains 1m ratings from 6k users and 4k items\final{, Netflix\footnote{\url{https://www.kaggle.com/netflix-inc/netflix-prize-data/}} contains 607k ratings from 10k users and 5k items,} while \final{the} Yahoo! dataset contains 300K ratings given by 15.4k users for 1k items.
\final{Furthermore, the Coat shopping dataset simulates the shopping for a coat in an online store. It contains about 7k ratings given by 290 users for 300 items.}
\final{All} datasets were collected from an unknown deployed recommendation system in a closed loop scenario.
In addition, \final{the} Yahoo! dataset has a unique feature: a subset of users (\final{5.4k}) were asked to rate 10 {\em randomly} selected items (open loop scenario). \final{Therefore, for a subset of users (5.4k), we have both \final{the} closed loop and open loop (random) feedback.} 
\final{Similarly, for the Coat dataset, each user \final{was} asked to provide ratings for 24 self-selected (closed loop) items and 16 randomly picked (open loop) items.}
\final{However, \final{in a real-world practical}} evaluation setup, \final{the training of the models \final{should} be} based on the \final{collected} closed loop feedback, \final{while the actual evaluation \final{could} be} based on \final{the} open loop (random) feedback \final{\final{because of} the \final{difficulties} of collecting open loop feedback in \final{the} real-world}, \final{as discussed in Section~1}.
\final{We} use the \final{normalised} Discounted Cumulative Gain (nDCG@k) metric for different cut-offs ($k=\{5, 10, 20, 30, 100\}$)\final{, while \final{we use} nDCG, without a rank cutoff, \final{to denote the case where the metric is computed on the ranking of} all items for each user.} 
\final{The user's rating value for \final{all} datasets is an integer $r\in[0,5]$. Our objective is to rank the most relevant items for each user. \final{Hence,} we binarise all explicit rating values by keeping the highly recommended items ($r\geq4$).}
\final{In the closed loop evaluation of the examined models, we use 80\% of the closed loop dataset (MovieLens, Yahoo! and Coat) for training and a 20\% random split for testing, \final{as is typical in RecSys evaluation~\cite{recsys_offline_options20}}. On the other hand, for the open loop evaluation (Yahoo! and Coat), we evaluate the trained models on all the provided open loop (randomised) feedback.}

\subsection{Recommendation Models}\label{sec:Recommendation Models}
\final{In our experiments, we use the implementations provided by the Cornac framework\footnote{\url{https://github.com/PreferredAI/cornac}} to evaluate the following models\footnote{\final{For ease of reference and compatibility, we use the same acronyms as used by the Cornac framework to denote the baselines.}}}:
\begin{itemize}
    \item \textbf{BO}: a simple baseline model that recommends a \final{random permutation of all available items to each user, regardless of their preferences.}
    \item \textbf{GA}: a baseline model that \final{recommends} the global average rating to each and every user, regardless of their preferences.
    \item \textbf{POP}: a baseline model that \final{\final{ranks} items based on their popularity, i.e.\ the number of times a specific item \final{is} rated, and suggests \final{these items} to \final{every} user regardless of their preferences.}
    \item \textbf{MF}: Matrix Factorisation~\cite{mf_netflix09} is a rating prediction model that represents both users and items as latent vectors. This model is optimised to predict the observed rating for each user-item pair.
    \item \textbf{PMF}: Probabilistic Matrix Factorisation~\cite{pmf07} is an extension of Matrix Factorisation~\cite{mf_netflix09} for large, sparse and imbalanced datasets. \final{There are two variants of this algorithm depending on how \final{the} users' preferences \final{are modelled}, i.e.\ linear and non-linear.} 
    We use both linear and non-linear variants in our experiments.
    \item \textbf{SVD++}: Singular Value Decomposition~\cite{svd_plus_plus08} is another rating prediction model \final{that} factorises the user-item rating matrix to a product of two lower rank matrices, i.e. user factors and item factors.
    \item \textbf{WMF}: Weighted Matrix Factorisation~\cite{als08} is another extension of Matrix Factorisation~\cite{mf_netflix09} that models confidence and uncertainty in addition to preferences.
    \item \textbf{NMF}: Non-negative Matrix Factorisation~\cite{nmf01} factorises the user-item rating matrix into user and item latent matrices, with the property that all matrices have no negative elements.
    \item \textbf{BPR}: Bayesian Personalised Ranking~\cite{bpr09} is a pair-wise ranking model. The BPR model is trained based on uniform negative sampling, i.e. we randomly sample items not interacted with as negative instances for each user. We use both normal and weighted variants in our experiments.
    \item \textbf{MMMF}: Maximum Margin Matrix Factorisation~\cite{mmmf08} is a type of matrix factorisation model \final{that is} optimised for \final{a} soft margin (Hinge) ranking loss.
    \item \textbf{NCF}: Neural Collaborative Filtering~\cite{neural_collaborative_filtering17} is a general neural network framework for collaborative filtering. We use Multi-Layer Perceptron (\textbf{MLP}), Generalised Matrix Factorisation (\textbf{GMF}) and Neural Matrix Factorisation (\textbf{NeuMF}) in our experiments.
\end{itemize}

We are interested \final{in} the {\em relative} order of \final{the models' performances} within closed loop and open loop \final{scenarios}.
\final{Therefore, }following previous research~\cite{offline_eval_spotify19}, \final{we vary the hyperparameter controlling the size of the latent factors} in \{10, 20, \ldots, 100\}, \final{for each model that supports this hyperparameter}\footnote{\final{All the examined models support a latent variable hyperparameter except BO, GA, POP and MLP.}}. \final{\final{This} leads} to \final{a total of} 104 models \final{that are evaluated} in our experiments. \final{Each} examined model differs on the algorithm or the latent variable size. \final{When discussing a particular model, we denote the hyperparameter value} in superscript - e.g.\ MF\textsuperscript{40} represents \final{the} MF model \final{using latent factor size} $m=40$.
For reproducibility, our code and the used datasets are available from: \url{https://github.com/terrierteam/stratified_recsys_eval}.

\subsection{Estimating Propensity Scores}
\label{sec:Estimating Propensity Scores}

The propensity score $p_{u,i}$ \final{is defined as} the tendency of the deployed model to expose item $i$ to user $u$ (Definition~\ref{def:propensity_score}). 
\final{Since it is not \final{realistic} for the deployed model to expose each and every item to the user, we have to estimate \final{the} propensity scores $p_{u,i}$ for various user and item pairs.}
Yang et al.~\cite{unbiased_offline_rec_eval18} proposed a simple method to estimate \final{the} propensity scores based on the following simplified assumption:
\begin{ass}
    The propensity score is user independent, i.e.\ $p_{u,i}=p_{*,i}$. This assumption was made to address the lack of auxiliary user information in public datasets.
\end{ass}
The user independent propensity score $p_{*,i}$ can be estimated \final{using a} two-step generative process~\cite{unbiased_offline_rec_eval18}:
\begin{equation}
    p_{*,i}=p_{*,i}^{select} * p_{*,i}^{interact|select}
\end{equation}
where $p_{*,i}^{select}$ is the prior probability that item $i$ is recommended by the deployed model and $p_{*,i}^{interact|select}$ is the conditional probability that the user interacts with item $i$ given that it is recommended.
Based on a set of mild assumptions, we can estimate the user independent propensity score $p_{*,i}$ as follows~\cite{unbiased_offline_rec_eval18}:
\begin{equation}
\hat{p}_{*, i} \propto\left(n_{i}^{*}\right)^{\left(\frac{\gamma+1}{2}\right)}
\label{eq:estimated_props}
\end{equation}
\looseness -1 where $n_{i}^{*}$ is the total number of times item $i$ is interacted with and $\gamma$ is a parameter that affects the propensity distributions over items with different observed popularity. 
\final{The power-law parameter $\gamma$ affects the propensity distributions over items and depends on the examined dataset.}
\final{Following previous research~\cite{powerlaw14}, }
we estimate the $\gamma$ parameter \final{using} maximum likelihood for each dataset.

\section{Experimental Results and Analysis}
\label{sec:Experimental Results and Analysis}

\final{We proposed a propensity-based stratified evaluation method in Section~\ref{sec:Propensity-based Stratified Evaluation} that takes into account the confounding role of the deployed model in the offline evaluation of recommendation systems.
In Section~\ref{sec:Experimental Setup}, we \final{proposed to} use Kendall's $\tau$ correlation coefficient to quantify the similarity between the relative order of many examined recommendation models in the proposed propensity-based stratified evaluation method \final{as well as in} the open loop (randomised) evaluation.}
\final{In the following, we conduct experiments with respect to the two research questions stated in Section~\ref{sec:Propensity-based Stratified Evaluation}, concerning the existence of \final{the} Simpson's paradox in the standard offline evaluation (Section~\ref{sec:Investigating Simpson's Paradox}) and the effectiveness of the proposed propensity-based stratified evaluation method (Section~\ref{sec:Evaluating Propensity-based stratified evaluation}).}

\subsection{RQ~\ref{rq:1}: Investigating Simpson's Paradox}
\label{sec:Investigating Simpson's Paradox}

\begin{table}
    \centering
    \caption{\final{An instance of} Simpson's paradox in the offline evaluation of recommender systems \final{using} nDCG on \final{the} MovieLens dataset~\cite{movielens15}. * denotes a significant difference compared to the other model (paired t-test, p <0.05).}
    \label{tbl:movielens}
    \begin{tabular}{l|c|c|c|c}
    Evaluation Method & BPR\textsuperscript{10} & WMF\textsuperscript{10} & Number of Ratings & Number of Items \\ \hline
    Holdout & \textbf{0.373}* & 0.369\phantom{*} & 200,018 & 3,449 \\ \hline
    IPS & \textbf{0.207}* & 0.204\phantom{*} & 200,018 & 3,449 \\ \hline
    Stratified & 0.344\phantom{*} & \textbf{0.351}* & 200,018 & 3,449 \\ 
    \multicolumn{1}{r|}{Q1} & 0.339\phantom{*} & \textbf{0.350}* & \textbf{197,600 (99\%)} & \textbf{3,445}  \\
    \multicolumn{1}{r|}{Q2} & \textbf{0.695}* & 0.418\phantom{*} & 2,418 (1\%) & 4 \\ 
    \end{tabular}
\end{table}

\begin{table}
    \centering
    \caption{\final{An instance of Simpson's paradox in the offline evaluation of recommender systems \final{using} nDCG on the Netflix dataset~\cite{mf_netflix09}. * denotes a significant difference compared to the other model (paired t-test, p <0.05).}}
    \label{tbl:netflix}
    \begin{tabular}{l|c|c|c|c}
    \final{Evaluation Method} & \final{MMMF\textsuperscript{100}} & \final{MF\textsuperscript{50}} & \final{Number of Ratings} & \final{Number of Items} \\ \hline
    \final{Holdout} & \final{\textbf{0.269}*} & \final{0.263\phantom{*}} & \final{121,460} & \final{3,962} \\ \hline
    \final{IPS} & \final{\textbf{0.246}\phantom{*}} & \final{0.245\phantom{*}} & \final{121,460} & \final{3,962} \\ \hline
    \final{Stratified} & \final{0.239\phantom{*}} & \final{\textbf{0.241}*} & \final{121,460} & \final{3,962} \\
    \multicolumn{1}{r|}{\final{Q1}} & \final{0.236\phantom{*}} & \final{\textbf{0.242}*} & \final{\textbf{109,452 (90\%)}} & \final{\textbf{3,944}}  \\
    \multicolumn{1}{r|}{\final{Q2}} & \final{\textbf{0.271}*} & \final{0.229\phantom{*}} & \final{12,008 (10\%)} & \final{18} \\ 
    \end{tabular}
\end{table}

Research Question~\ref{rq:1} \final{investigates} the confounding effect of the deployed model on the offline evaluation of recommendation systems based on closed loop feedback datasets as mentioned in Hypothesis~\ref{hyp:closed_loop}. In Section~\ref{sec:Estimating Propensity Scores}, we \final{presented a simple statistical method to} \final{represent} the deployed model's \final{characteristics} as propensity scores. In the following, we use the estimated propensity scores $\hat{p}_{*, i}$ to partition the examined dataset into two equal size strata, namely \final{the} Q1 and Q2 strata. Since we \final{estimate} the propensity scores based on the total number of times each item is interacted with \final{according to Equation~\eqref{eq:estimated_props}}, the Q1 and Q2 \final{strata represent} the users' interactions with the long-tail and head items, respectively.

\final{First}, \final{we highlight some instances of the Simpson's paradox in both the examined closed loop (MovieLens \final{and Netflix}) and open loop (Yahoo! and Coat) datasets.}
Tables~\ref{tbl:movielens} \final{\& \ref{tbl:netflix}} compare the effectiveness of the examined evaluation methods (namely \final{the} holdout, IPS and \final{our} proposed stratified evaluation \final{methods}) on \final{the MovieLens \final{and Netflix datasets}.}
For the sake of simplicity, \final{in the following,} we focus on \final{the Movielens dataset, analysing} two models (BPR\textsuperscript{10}, WMF\textsuperscript{10}) and one metric (nDCG)\footnote{In Section~\ref{sec:Evaluating Propensity-based stratified evaluation}, we compare the effectiveness of all the examined \final{models} for different nDCG cut-offs based on \final{the} Kendall's $\tau$ rank correlation coefficient.}. 
We observe that BPR\textsuperscript{10} is significantly better than WMF\textsuperscript{10} \final{using} both the holdout and counterfactual (IPS) evaluation methods. 
However, \final{the} stratified analysis on the same test dataset (Q1 and Q2 \final{strata}) reveals that WMF\textsuperscript{10} is significantly better than BPR\textsuperscript{10} for \final{the} Q1 stratum while BPR\textsuperscript{10} is the dominant model for the Q2 stratum. 
In addition, we observe that \final{the} Q1 and Q2 \final{strata cover} 99\% and 1\% of \final{the} user-item interactions, respectively. \final{In fact}, WMF\textsuperscript{10} is significantly better than BPR\textsuperscript{10} for 99\% of the examined test dataset (i.e.\ the Q1 stratum) but this trend reverses when we \final{combine} \final{it with the} 1\% feedback in the Q2 stratum\final{, as represented in the \final{h}oldout evaluation}. Therefore, the \final{real} question is {\em \final{whether} BPR\textsuperscript{10} \final{does} perform better than WMF\textsuperscript{10} as \final{recognised} by the holdout evaluation method?} 
The stratified analysis reveals that WMF\textsuperscript{10} should be recognised as the superior model \final{by} any reasonable evaluation method as \final{recognised} by \final{our} proposed stratified evaluation. We \final{note} that BPR\textsuperscript{10} is the superior model only for 1\% of the feedback dataset and both the holdout and IPS \final{evaluation methods} are influenced \final{by} a {\em \final{very small} minority} of user-item interactions in the test dataset, i.e.\ 1\% of user-item interactions \final{in the Q2 stratum}.
\final{This stratum \final{corresponds} } to only 4 items among 3,499 items in the MovieLens dataset. \final{The same pattern \final{is also observed for} the Netflix dataset between \final{the} MMMF\textsuperscript{100} and MF\textsuperscript{50} models \final{presented} in Table~\ref{tbl:netflix}.}

\begin{table}
    \centering
    \caption{\final{An instance of} Simpson's paradox in the offline evaluation of recommender systems \final{using} nDCG on \final{the} Yahoo! dataset~\cite{collaborative_ranking_non_random_missing09}. * denotes a significant difference compared to the other model (paired t-test, p <0.05).}
    \label{tbl:yahoo}
    \begin{tabular}{l|c|c|c|c}
    Evaluation Method & BPR\textsuperscript{40} & MF\textsuperscript{40} & Number of Ratings & Number of Items \\ \hline
    Holdout  & \textbf{0.270}* & 0.266\phantom{*} & 62,341 & 1,000 \\ \hline
    IPS  & 0.238\phantom{*} & \textbf{0.239}\phantom{*} & 62,341 & 1,000 \\ \hline
    Stratified  & 0.249\phantom{*} & \textbf{0.256}* & 62,341 & 1,000 \\
    \multicolumn{1}{r|}{Q1} & 0.218\phantom{*} & \textbf{0.250}* & \textbf{57,467 (92\%)} & \textbf{995}  \\
    \multicolumn{1}{r|}{Q2} & \textbf{0.622}* & 0.330\phantom{*} & 4,874  (8\%)\phantom{*} & 5 \\ \hline\hline
    Open Loop & 0.169\phantom{*} & \textbf{0.172}* & 54,000 & 1,000 \\ 
    \end{tabular}
\end{table}

\begin{table}
    \centering
    \caption{\final{An instance of} Simpson's paradox in the offline evaluation of recommender systems \final{using} nDCG on \final{the} Coat dataset~\cite{rec_treatment16}. * denotes a significant difference compared to the other model (paired t-test, p <0.05).}
    \label{tbl:coat}
    \begin{tabular}{l|c|c|c|c}
    Evaluation Method & GMF\textsuperscript{20} & SVD\textsuperscript{20} & Number of Ratings & Number of Items \\ \hline
    Holdout  & \textbf{0.245}\phantom{*} & 0.244\phantom{*} & 1,392 & 286 \\ \hline
    IPS  & 0.235\phantom{*} & \textbf{0.237}\phantom{*} & 1,392 & 286 \\ \hline
    Stratified  & 0.227\phantom{*} & \textbf{0.230}\phantom{*} & 1,392 & 286 \\
    \multicolumn{1}{r|}{Q1} & 0.199\phantom{*} & \textbf{0.215}* & \textbf{1,301 (93\%)} & \textbf{281}  \\
    \multicolumn{1}{r|}{Q2} & \textbf{0.612}* & 0.434\phantom{*} & \phantom{**}91 (7\%)\phantom{*} & 5 \\ \hline\hline
    Open Loop & 0.248\phantom{*} & \textbf{0.264}* & 1,392 & 286 \\ 
    \end{tabular}
\end{table}

\begin{figure}
    \centering
    \begin{subfigure}[b]{\textwidth}
        \includegraphics*[width=\linewidth]{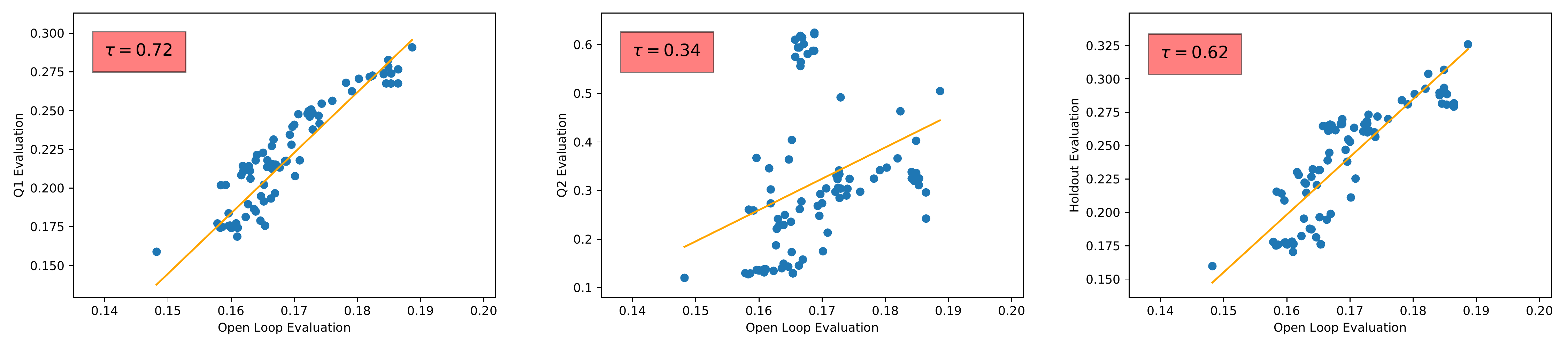}
        \caption{Yahoo! Dataset}
        \label{fig:yahoo_vis}
    \end{subfigure}
    \begin{subfigure}[b]{\textwidth}
        \includegraphics*[width=\linewidth]{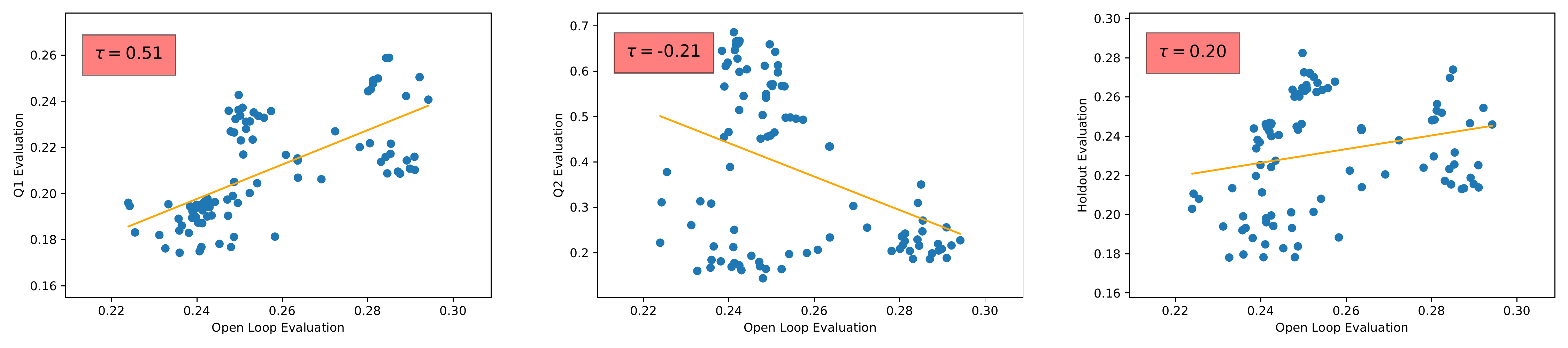}
        \caption{Coat Dataset}
        \label{fig:coats_vis}
    \end{subfigure}
    \caption{\final{The correlation of the examined models between the open loop evaluation (x axis) and the specified closed loop evaluation (y axis) in the Yahoo! dataset (a) and the Coat dataset (b). All Kendall's $\tau$ correlations are statistically significant \final{according to} Steiger's method~\cite{steiger80} (p <0.05). The orange line represents the best fit for the open/closed loop data based on a linear regression model.}}
    \label{fig:simpson_vis}
\end{figure}

\looseness -1 In the MovieLens \final{and Netflix} dataset\final{s}, we do not \final{have} access to open loop feedback, i.e.\ feedback that \final{was} collected from \final{a} randomised exposure \final{to items}. As a result, we \final{cannot} measure the actual \final{performances} of the models in an open loop scenario in comparison to the corresponding closed loop scenario. As mentioned in Section~\ref{sec:Datasets and Evaluation metrics}, we have \final{some} open loop feedback for \final{all the users in the Coat dataset and} a fraction of users in \final{the} Yahoo! dataset. \final{Tables~\ref{tbl:yahoo} and~\ref{tbl:coat} \final{present} performance comparisons of two pairs of \final{models} on the Yahoo! and Coat datasets. In particular, in Table~\ref{tbl:yahoo}, we compare BPR\textsuperscript{40} and MF\textsuperscript{40} \final{using} the nDCG evaluation metric, and observe} that BPR\textsuperscript{40} is significantly better than MF\textsuperscript{40} \final{using} the classical holdout evaluation method. On the other hand, the IPS evaluation method \final{favours} MF\textsuperscript{40} but the difference is not significant based on \final{the} paired t-test ($p <0.05$).
However, \final{the} stratified analysis based on the estimated propensities (\final{the} Q1 and Q2 \final{strata}) reveals that MF\textsuperscript{40} is significantly better than BPR\textsuperscript{40} for \final{the} Q1 stratum, which covers 92\% of feedback in the examined test dataset while BPR\textsuperscript{40} is the dominant model for the Q2 stratum, which covers only 8\% of feedback in the closed loop test dataset. 
Indeed, MF\textsuperscript{40} significantly
\final{outperforms} BPR\textsuperscript{40} for the {\em majority} of feedback and items (92\% and 99.5\%, respectively) in the examined test dataset.
Therefore, \final{we argue that} \final{MF\textsuperscript{40}} should be recognised as a better model by any reasonable evaluation method as \final{recognised} by the proposed stratified evaluation method and the open loop evaluation where random items were exposed to the users. 
When we evaluate the models based on the aggregation of \final{the} Q1 and Q2 \final{strata} (i.e.\ holdout evaluation), a \final{small} minority of 5 items out of 1,000 total items \final{corresponding} to only 8\% of the total user-item interactions in \final{the} Yahoo! dataset, \final{plays} a confounding factor in the offline evaluation of the examined models. 
\final{When considering the Coat dataset (Table~\ref{tbl:coat}), we also observe the same phenomenon when assessing the effectiveness of \final{the} GMF\textsuperscript{20} and SVD\textsuperscript{20} recommendation models.} The observed phenomenon in Table~\ref{tbl:movielens}, Table~\ref{tbl:yahoo} \final{and Table~\ref{tbl:coat}} can be explained by the fact that \final{all the examined datasets (MovieLens, \final{Netflix,} Yahoo! and Coat)} were collected from a closed loop scenario. Therefore, the collected closed loop dataset is skewed towards the deployed model's \final{characteristics} (as captured in the Q2 stratum \final{in Table~\ref{tbl:movielens}, Table~\ref{tbl:yahoo} \final{and Table~\ref{tbl:coat}}}).

\final{Next, to fully answer Research Question 1, we assess the generalisation of the above observations, by verifying the prevalence of the Simpson's paradox in the evaluation of all our 104 models}. \final{Figure~\ref{fig:simpson_vis} shows the correlation of all the examined models (104 models as described in Section~\ref{sec:Datasets and Evaluation metrics}) between the open loop evaluation and the specified closed loop evaluation for both \final{the} Yahoo! and Coat datasets.} 
\final{We observe that \final{the} models exhibit \final{an imbalanced} performance on the Q1 and Q2 strata, \final{i.e.\ the nDCG scores for the Q1 stratum \final{are not} in proportion to the Q2 stratum ($nDCG_{Q2} \gg nDCG_{Q1}$)}. 
On the other hand, the Kendall's $\tau$ correlation \final{between the Q1 stratum and the open loop evaluation is much higher compared to the corresponding} Q2 stratum ($\tau_{Q1} \gg \tau_{Q2}$).}
\final{Specifically, for the Coat dataset (Figure~\ref{fig:coats_vis}), the closed loop evaluation based on the Q2 stratum has a {\em negative} correlation with the open loop evaluation.}
\final{As a result, combining these two {\em heterogeneous} strata in the holdout evaluation -- as shown in Figure~\ref{fig:yahoo_vis} and Figure~\ref{fig:coats_vis} -- without taking into account that the Q1 stratum covers the majority of feedback (92\% and 93\% of the total feedback in the \final{Yahoo! and Coat datasets}, respectively), leads to \final{unexpected} consequences\footnote{This is in contrast to the Q2 stratum that covers only the remaining feedback, which is a small fraction of the dataset.}. \final{In particular,} the Kendall's $\tau$ correlation between the open loop evaluation and the holdout evaluation is significantly lower than the corresponding correlation between the Q1 stratum and the open loop evaluation, i.e.\ $0.62 < 0.72$ and $0.20 < 0.51$ for the Yahoo! and the Coat datasets, respectively.}
\final{This answers} Research Question~\ref{rq:1}: the offline \final{holdout} evaluation of recommendation systems is influenced by the deployed model's \final{characteristics} \final{(as captured in the Q2 stratum)}, leading to \final{a} conclusion \final{prone to} the Simpson's paradox. 

\final{\final{To summarise, in} this section, we highlighted the existence of \final{the} Simpson's paradox in the standard offline evaluation of recommendation systems based on both closed loop (MovieLens \final{and Netflix}) and open loop (Yahoo! and Coat) datasets.}
In the following, we investigate the effectiveness of \final{our} proposed \final{propensity-based} stratified evaluation \final{in addressing the issue}.

\subsection{RQ~\ref{rq:2}: \final{Assessing} \final{the Proposed} Propensity-based Stratified \final{Evaluation} Method}
\label{sec:Evaluating Propensity-based stratified evaluation} 

\looseness -1 In this section, we investigate \final{\final{the extent to which} the proposed propensity-based stratified evaluation method (Section~\ref{sec:Propensity-based Stratified Evaluation}) \final{leads to conclusions that are aligned} with the \final{results obtained} from the open loop (randomised) evaluation, compared with} both the counterfactual (IPS) and the holdout evaluation methods \final{(Section~\ref{sec: Offline Evaluation Methods})}. \final{Indeed, our} \final{objective} is to investigate to what extent each evaluation method correlates with the \final{open loop evaluation}. \final{In contrast, we consider  open loop evaluation to be more akin to online evaluation (A/B testing or interleaving) where the evaluation process is not influenced by any confounders. }
\final{As mentioned in Section~\ref{sec:Recommendation Models}, we} \final{make use of the} Kendall's $\tau$ rank correlation coefficient between the examined evaluation method and the open loop evaluation for all the examined recommendation models. 

\begin{table}
    \centering
    \caption{Kendall's $\tau$ rank correlation coefficient between the examined evaluation methods and \final{the} {\em open loop (randomised)} evaluation on \final{the} Yahoo!~\cite{collaborative_ranking_non_random_missing09} and Coat~\cite{rec_treatment16} datasets. \final{\textsuperscript{1/2/3/4/5}} denote a significant difference compared to the \final{indicated evaluation method} (Steiger's method~\cite{steiger80}, p <0.05).}
    \label{tbl:correlation}
    \begin{tabular}{c|l|c|c|c|c|c}
    Dataset & Metric & Holdout\final{\textsuperscript{1}} & IPS\final{\textsuperscript{2}} & Stratified\final{\textsuperscript{3}} & Q1\final{\textsuperscript{4}} & Q2\final{\textsuperscript{5}} \\\hline
    \multirow{6}{*}{\rotatebox[origin=c]{90}{Yahoo!}} & nDCG@5   & \phantom{*}0.599\final{\textsuperscript{3/4/5\phantom{/0}}} & 0.641\final{\textsuperscript{4/5\phantom{/0}\phantom{/0}}} & \textbf{0.700}\final{\textsuperscript{1/4/5\phantom{/0}}} & 0.407\final{\textsuperscript{1/2/3\phantom{/0}}} & \phantom{*}0.414\final{\textsuperscript{1/2/3\phantom{/0}}}  \\
    & nDCG@10  & \phantom{*}0.693\final{\textsuperscript{4/5\phantom{/0}\phantom{/0}}} & 0.704\final{\textsuperscript{4/5\phantom{/0}\phantom{/0}}} & 0.697\final{\textsuperscript{4/5\phantom{/0}\phantom{/0}}} & 0.559\final{\textsuperscript{1/2/3\phantom{/0}}} & \phantom{*}0.501\final{\textsuperscript{1/2/3\phantom{/0}}}  \\
    & nDCG@20  & \phantom{*}0.729\final{\textsuperscript{4/5\phantom{/0}\phantom{/0}}} & 0.721\final{\textsuperscript{4/5\phantom{/0}\phantom{/0}}} & 0.716\final{\textsuperscript{4/5\phantom{/0}\phantom{/0}}} & 0.636\final{\textsuperscript{1/2/3\phantom{/0}}} & \phantom{*}0.523\final{\textsuperscript{1/2/3\phantom{/0}}}  \\
    & nDCG@30  & \phantom{*}0.652\final{\textsuperscript{3/5\phantom{/0}\phantom{/0}}} & 0.653\final{\textsuperscript{3/5\phantom{/0}\phantom{/0}}} & \textbf{0.713}\final{\textsuperscript{1/2/5\phantom{/0}}} & 0.673\final{\textsuperscript{5\phantom{/0}\phantom{/0}\phantom{/0}}} & \phantom{*}0.450\final{\textsuperscript{1/2/3/4}}  \\
    & nDCG@100 & \phantom{*}0.701\final{\textsuperscript{3/4/5\phantom{/0}}} & 0.724\final{\textsuperscript{3/5\phantom{/0}\phantom{/0}}} & \textbf{0.774}\final{\textsuperscript{1/2/5\phantom{/0}}} & 0.773\final{\textsuperscript{1/5\phantom{/0}\phantom{/0}}} & \phantom{*}0.348\final{\textsuperscript{1/2/3/4}}  \\
    & nDCG     & \phantom{*}0.622\final{\textsuperscript{3/4/5\phantom{/0}}} & 0.644\final{\textsuperscript{3/4/5\phantom{/0}}} & 0.710\final{\textsuperscript{1/2/5\phantom{/0}}} & \textbf{0.722}\final{\textsuperscript{1/2/5\phantom{/0}}} & \phantom{*}0.336\final{\textsuperscript{1/2/3/4}}  \\ \hline
    
    \multirow{6}{*}{\rotatebox[origin=c]{90}{Coat}} & nDCG@5 & -0.040\textsuperscript{3/4/5\phantom{/0}} & 0.051\textsuperscript{3/4/5\phantom{/0}} & 0.116\textsuperscript{1/2/4/5} & \textbf{0.362}\textsuperscript{1/2/3/5} & -0.288\textsuperscript{1/2/3/4} \\
     & nDCG@10 & \phantom{*}0.121\textsuperscript{3/4/5\phantom{/0}} & 0.165\textsuperscript{3/4/5\phantom{/0}} & 0.219\textsuperscript{1/2/4/5} & \textbf{0.418}\textsuperscript{1/2/3/5} & -0.240\textsuperscript{1/2/3/4} \\
     & nDCG@20 & \phantom{*}0.126\textsuperscript{4/5\phantom{/0}\phantom{/0}} & 0.118\textsuperscript{4/5\phantom{/0}\phantom{/0}} & 0.172\textsuperscript{4/5\phantom{/0}\phantom{/0}} & \textbf{0.450}\textsuperscript{1/2/3/5} & -0.195\textsuperscript{1/2/3/4} \\
     & nDCG@30 & \phantom{*}0.165\textsuperscript{4/5\phantom{/0}\phantom{/0}} & 0.159\textsuperscript{4/5\phantom{/0}\phantom{/0}} & 0.213\textsuperscript{4/5\phantom{/0}\phantom{/0}} & \textbf{0.467}\textsuperscript{1/2/3/5} & -0.207\textsuperscript{1/2/3/4} \\
     & nDCG@100 & \phantom{*}0.327\textsuperscript{3/4/5\phantom{/0}} & 0.360\textsuperscript{3/4/5\phantom{/0}} & 0.393\textsuperscript{1/2/4/5} & \textbf{0.547}\textsuperscript{1/2/3/5} & -0.186\textsuperscript{1/2/3/4} \\
     & nDCG & \phantom{*}0.202\textsuperscript{3/4/5\phantom{/0}} & 0.225\textsuperscript{3/4/5\phantom{/0}} & 0.283\textsuperscript{1/2/4/5} & \textbf{0.509}\textsuperscript{1/2/3/5} & -0.208\textsuperscript{1/2/3/4} \\
    \end{tabular}
\end{table}

Table~\ref{tbl:correlation} \final{shows the} Kendall's $\tau$ rank correlation coefficient between the examined evaluation methods and the ground truth, i.e.\ open loop (randomised) evaluation.  
\final{On analysing the table, we observe that the
$\tau$ correlation between the holdout evaluation and the open loop evaluation is medium-to-strong ($0.599\leq \tau \leq 0.729$) for the Yahoo! dataset, and weaker (\final{and $-0.40 \leq \tau \leq 0.327$) for the Coat dataset}, with correlations \final{on both datasets} rising slightly for larger cut-offs.} 
\final{Indeed, we note that the Coat dataset used a different data generation process compared to the MovieLens\final{, Netflix and Yahoo! datasets}~\cite{rec_treatment16} (\final{i.e.} making use of crowdsourcing users rather than real users). In addition, the number of examined users and items are lower than \final{those in} the Yahoo! dataset.}
\final{This finding \final{agrees} with} previous research~\cite{offline_eval_spotify19,contrasting_offline_online_recsys16} \final{who found} \final{that the} \final{offline evaluation based on standard rank-based metrics}
does not \final{necessarily} correlate with \final{the} online evaluation (A/B testing or interleaving) results and researchers have questioned \final{the} validity of offline evaluation~\cite{recsys_offline_options20,ir_metrics_robustness_recsys18,recsys_eval19}.

\final{Next, we consider the Kendall's $\tau$ correlation between the counterfactual evaluation (IPS) and the open loop evaluation methods. \final{Table~\ref{tbl:correlation}} reveals that the IPS evaluation method \final{performs slightly} better than the holdout evaluation, i.e.\ the correlation values for IPS are greater than the holdout evaluation method for \final{the majority of nDCG cut-offs in both datasets}. However, none of these \final{correlation} differences are statistically significant according to \final{the} Steiger's test.} 
\final{Therefore, we \final{conclude} that \final{the} counterfactual (IPS) evaluation method \final{does not better represent unbiased user preferences during \final{the} evaluation than \final{the} holdout evaluation}} \final{for the examined datasets.} \final{This is because} in the IPS evaluation method, each feedback is \final{weighted} {\em individually} based on its propensity. The feedback sampling process, i.e.\ the process \final{from which the} \final{closed loop} feedback \final{was} collected, is not a {\em random} process. As a result, the sampled feedback \final{is} skewed and the corresponding estimated propensity scores are imbalanced. 
\final{In this situation,} reweighting \final{the} feedback based on \final{its} disproportionate propensities leads to high variance \final{ in the} \final{models' performance estimation} \final{using} the IPS evaluation method~\cite{batch_logged_bandit_counterfactual15,unbiased_offline_rec_eval18}.

\final{We now consider the correlations exhibited by the proposed stratified evaluation in Table~\ref{tbl:correlation}.} \final{This} reveals that the stratified evaluation performs significantly better than both the holdout \final{and} IPS evaluation methods\footnote{\final{The holdout and IPS significance tests are represented as \textsuperscript{1/2}, respectively in the Stratified column of Table~\ref{tbl:correlation}.}} \final{for the majority of nDCG cut-offs in both datasets, specifically \final{at} deeper cut-offs} \final{(Steiger's significance test~\cite{steiger80}, $p<0.05$)}. 
\final{For example, considering nDCG as the evaluation metric, we observe that the proposed stratified evaluation method performs \final{significantly} better than both the holdout evaluation and the IPS evaluation, i.e.\ $0.710 > 0.622$ and $0.710 > 0.644$ for the Yahoo! dataset, while $0.283 > 0.202$ and $0.283 > 0.225$ for the Coat dataset.}
\final{\final{Overall,} we find that \final{our} propensity-based stratified evaluation \final{method} performs significantly better than both the holdout \final{and} counterfactual evaluation \final{methods} \final{for deeper nDCG cut-offs in both datasets}, which answers Research Question~\ref{rq:2}.}

\final{However,} \final{the \final{performances} of \final{the holdout evaluation, the IPS evaluation and the proposed propensity-based stratified evaluation method \final{are} not significantly different} for shallow nDCG cut-offs (\final{$10\leq k \leq20$ and $20\leq k \leq30$ for the Yahoo! dataset and the Coat dataset, respectively)}} \final{while}
\final{the observed correlations generally increase for deeper nDCG cut-offs.} 
\final{The \final{increased} sensitivity for deeper rank cut-offs} supports previous research~\cite{ir_metrics_robustness_recsys18} on the greater robustness to sparsity and \final{the} discriminative power of deeper cut-offs on \final{the evaluation of recommendation systems}. Valcarce et al.~\cite{ir_metrics_robustness_recsys18} found that since only a few items are exposed to the user during the feedback sampling process, the use of deeper cut-offs allow researchers to perform more robust and discriminative evaluations of recommender systems. \final{Similar observations were made by \citet{10.1007/s10791-012-9209-9} concerning the use of deeper rank cut-offs for training listwise learning to rank techniques for web search.}

\begin{figure}
    \centering
    \begin{subfigure}[b]{0.45\textwidth}
        \resizebox{\textwidth}{!}{%
            \begin{tikzpicture}
            	\begin{axis}[
                    ybar,
                    enlargelimits=0.15,
                    legend style={at={(0.5,-0.15)},
                    anchor=north,legend columns=-1},
                    nodes near coords,
                    nodes near coords align={vertical},
            		xlabel=\final{Number of strata},
            		ylabel=\final{Kendall's $\tau$},
            		xtick={1,2,3,4,5,6,7,8,9,10}]
            	\addplot coordinates {
            		(1, 0.202) %
            		(2, 0.282) %
            		(3, 0.235)
            		(4, 0.244)
            		(5, 0.285)
            		(6, 0.276)
            		(7, 0.268)
            		(8, 0.249)
            		(9, 0.300)
            		(10, 0.251)
            	};
                \addplot[red,sharp plot,dashed,nodes near coords={}] coordinates {
                        (2,0.266) 
                        (10,0.266)
                    };
                \node [above, red] at (axis cs:10,0.266) {$0.266 \pm 0.021$};
            	\node [above, red, font=\Large] at (axis cs: 2,0.007+0.282) {$\ast$};
            	\node [above, red, font=\Large] at (axis cs: 5,0.007+0.285) {$\ast$};
            	\node [above, red, font=\Large] at (axis cs: 6,0.007+0.276) {$\ast$};
            	\node [above, red, font=\Large] at (axis cs: 7,0.007+0.268) {$\ast$};
            	\node [above, red, font=\Large] at (axis cs: 9,0.007+0.300) {$\ast$};
            	\end{axis}
            \end{tikzpicture}
        }%
        \caption{\final{Coat Dataset}}
        \label{fig:coats_strata}
    \end{subfigure}\hfill
    \begin{subfigure}[b]{0.45\textwidth}
        \resizebox{\textwidth}{!}{%
            \begin{tikzpicture}
            	\begin{axis}[
                    ybar,
                    enlargelimits=0.15,
                    legend style={at={(0.5,-0.15)},
                    anchor=north,legend columns=-1},
                    nodes near coords,
                    nodes near coords align={vertical},
            		xlabel=\final{Number of strata},
            		ylabel=\final{Kendall's $\tau$},
            		xtick={1,2,3,4,5,6,7,8,9,10}]
            	\addplot coordinates {
            		(1, 0.622) %
            		(2, 0.710) %
            		(3, 0.722)
            		(4, 0.713)
            		(5, 0.718)
            		(6, 0.728)
            		(7, 0.720)
            		(8, 0.718)
            		(9, 0.683)
            		(10, 0.639)
            	}; %
                \addplot[red,sharp plot,dashed,nodes near coords={}] coordinates {
                        (2,0.703) 
                        (10,0.703)
                    };
            \node [above, red] at (axis cs:10,0.703) {$0.706 \pm 0.026$};
            \node [above, red, font=\Large] at (axis cs: 2,0.007+0.710) {$\ast$};
            \node [above, red, font=\Large] at (axis cs: 3,0.007+0.722) {$\ast$};
            \node [above, red, font=\Large] at (axis cs: 4,0.007+0.713) {$\ast$};
            \node [above, red, font=\Large] at (axis cs: 5,0.007+0.718) {$\ast$};
            \node [above, red, font=\Large] at (axis cs: 6,0.007+0.728) {$\ast$};
            \node [above, red, font=\Large] at (axis cs: 7,0.007+0.720) {$\ast$};
            \node [above, red, font=\Large] at (axis cs: 8,0.007+0.718) {$\ast$};
            \end{axis}
            \end{tikzpicture}
        }%
        \caption{\final{Yahoo! Dataset}}
        \label{fig:yahoo_strata}
    \end{subfigure}
    \caption{\final{The effect of the number of strata ($X$) on the performance of the proposed stratified evaluation. %
    The horizontal and vertical axes \final{correspond} to the number of strata and the correlation to the open loop (randomised) evaluation, respectively. $\ast$ denotes a significant difference compared to the \final{standard} holdout evaluation method ($X=1$) based on Steiger's method~\cite{steiger80}, $p <0.05$. The horizontal dashed line represents the mean correlations for $2 \leq X \leq 10$.}}
    \label{fig:strata}
\end{figure}
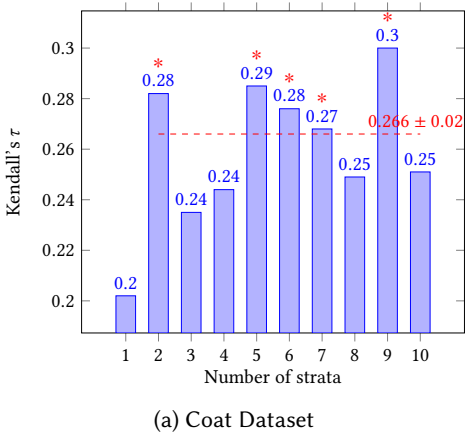
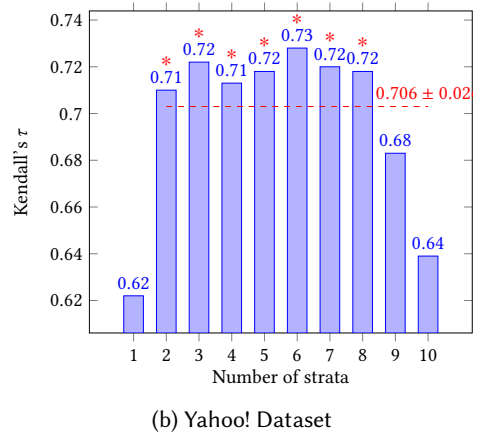

Further experiments on the feedback sub-populations (\final{the} Q1 and Q2 \final{strata} in Table~\ref{tbl:correlation}) reveal that the offline evaluation of the examined models, \final{which is} only based on \final{the} long-tail feedback (Q1 stratum) better correlates with the open loop evaluation. 
\final{Specifically, for the Coat dataset, the offline evaluation of the examined models based on the Q2 stratum has a {\em negative} correlation with the open loop evaluation.}
\final{This} supports previous research \final{such as} Cremonesi et al.~\cite{recsys_performance_topk10}, \final{who} found that the very few top popular items skew the rank-based evaluation of \final{the} recommendation systems. They suggested to exclude the extremely popular items while evaluating the effectiveness of recommendation models.

\final{\final{While Table~\ref{tbl:correlation} demonstrates} the effectiveness of the proposed stratified evaluation method for two sub-populations (Q1 and Q2 stratum), \final{we next examine the effect} of the number of strata used. \final{In particular},}
\final{Figure~\ref{fig:strata} \final{demonstrates} the effect of the number of strata in the proposed stratified evaluation for both \final{the Coat and Yahoo!} datasets. The horizontal axes ($X=\{1,2,...,10\}$) \final{represent} the number of strata while the vertical axes \final{represent} the correlation \final{between the proposed stratified evaluation method and} the open loop (randomised) evaluation \final{for all the 104 examined models}. When we have only one stratum (i.e.\ $X=1$ in Figure~\ref{fig:coats_strata} and Figure~\ref{fig:yahoo_strata}), the proposed stratified evaluation method \final{corresponds} to the holdout evaluation presented in Table~\ref{tbl:correlation}.
We observe that for all the examined number of strata ($2 \leq X \leq 10$) in both datasets, the proposed stratified evaluation \final{method} better correlates with the open loop (randomised) evaluation compared with the holdout evaluation (i.e. $X=1$). However, the number of strata has a marginal effect on the correlation between the proposed stratified evaluation and the open loop (randomised) evaluation, i.e. the mean correlations between the proposed stratified evaluation ($ 2 \leq X \leq 10$) for \final{the} Coats and Yahoo! datasets are $0.266 \pm 0.021$ and $0.706 \pm 0.026$, respectively, compared to $0.202$ and $0.622$ correlations in \final{the} holdout evaluation ($X=1$).
In addition, for $2 \leq X \leq 10$, the majority of cases (5 and 7 out of 9 for \final{the} Coats and Yahoo! datasets, respectively) demonstrate significantly higher correlations than the holdout evaluation.
\final{Note that the number of strata to use for each dataset can be determined using the stratification principle~\cite{stratified_sampling_ch11}, which favours the use of strata \final{such} that within each stratum, the users' feedback \final{is} as similar as possible.}}
\final{Although different datasets have different levels of closed loop effect depending on the deployed model's characteristics,}
our experiments reveal that \final{without further information about the deployed model,} stratifying closed loop feedback into sub-populations based on the estimated propensities 
allow researchers to \final{account for} the closed loop effect from the collected feedback dataset.

\section{Conclusions}
\label{sec:Conclusions}

The first key step in evaluating recommendation models based on feedback datasets is to understand how the feedback data was generated. 
As for any causal analysis, we started from a hypothesis about the data generating process (Hypothesis~\ref{hyp:closed_loop}).
We investigated the influence of the deployed recommendation model on the closed loop feedback data generation \final{process}. \final{Next, we studied the impact of this collected closed loop feedback data on the offline evaluation of a new recommendation model}.
Our in-depth experiments based on both the closed loop and open loop (randomised) feedback datasets \final{revealed} that the \final{standard} holdout evaluation \final{method suffers} from the Simpson's paradox, i.e.\ a \final{very small} minority of items, which are repeatedly exposed by the deployed model \final{plays} a confounding factor in the holdout evaluation where all feedback \final{is} aggregated together without accounting for the deployed \final{model's} confounding effect.
Consequently, our experiments \final{highlighted} that when evaluating a recommendation model in an offline fashion, the test set should be chosen carefully in order to free \final{it from any} confounders, \final{and} specifically \final{from the} closed loop algorithmic effect. Our findings \final{on \final{four} well-known closed loop and open loop (randomised) feedback datasets} provided a theoretical and experimental explanation of the  \final{concerns regarding the offline evaluation of recommendation systems} observed in the recent literature~\cite{closed_loop_feedback_sigir20,recsys_eval19, algorithmic_confounding18,unbiased_offline_rec_eval18,ir_metrics_robustness_recsys18}.
In addition, we proposed a novel propensity-based stratified evaluation \final{method}, which takes into account the role of the deployed model in the offline evaluation \final{to more accurately estimate the performance of a given new recommendation model}. The proposed model significantly \final{outperforms} the classical holdout evaluation and the counterfactual evaluation methods \final{for deeper nDCG cut-offs}, i.e.\ the proposed model better represents the actual relative order of the examined models \final{using an} open loop \final{setup}.
Our work aims \final{to bring attention to} the challenges of the offline evaluation based on closed loop feedback datasets and \final{to show how the application of} causal reasoning techniques, \final{through our proposed propensity-based stratified evaluation method}, \final{allows to counter} the confounding effect of the deployed model \final{so as to obtain a more accurate evaluation of a recommender system}.

\begin{acks}
\looseness -1 \final{This work has been supported by EPSRC grant EP/R018634/1: Closed-Loop Data Science for Complex, Computationally- \& Data-Intensive Analytics.}
\end{acks}

\bibliographystyle{ACM-Reference-Format}
\bibliography{references}

\end{document}